\documentclass[conference]{IEEEtran}

\usepackage{amsfonts,amsmath,amsthm}
\usepackage{graphicx,xspace,epsfig,syntonly,dsfont}
\usepackage{url,cite,bm,bbm}
\usepackage{algorithmicx}
\usepackage{balance}
\usepackage{subfigure}
\usepackage{stfloats}
\usepackage{diagbox}
\usepackage{multirow}
\newtheorem{remark}{Remark}

\long\def\symbolfootnote[#1]#2{\begingroup
\def\thefootnote{\fnsymbol{footnote}}
\footnote[#1]{#2}\endgroup}

\IEEEoverridecommandlockouts
\allowdisplaybreaks[4]

\title{\Large \bf Variation-cognizant Probabilistic Power Flow Analysis via Multi-task Learning}
\author{
    \IEEEauthorblockN{Kejun Chen and Yu Zhang} 
    \IEEEauthorblockA{Department of Electrical and Computer Engineering}
    \IEEEauthorblockA{University of California, Santa Cruz}
    \IEEEauthorblockA{Emails:\texttt\{kchen158,\,zhangy\}@ucsc.edu}
    \thanks{This work was supported in part by the Faculty Research Grant of UC Santa Cruz, Seed Fund Award from CITRIS and the Banatao Institute at the University of California, and the Hellman Fellowship.}
}

\begin{document}
\maketitle

\begin{abstract}
With an increasing high penetration of solar photovoltaic generation in electric power grids, voltage phasors and branch power flows experience more severe fluctuations. In this context, probabilistic power flow (PPF) study aims at characterizing the statistical properties of the state of the system with respect to the random power injections. To avoid repeated power flow calculations involved in PPF study, the present paper leverages regression algorithms and neural networks to improve the estimation performance and speed up the computation. Specifically, based on the variation level of the voltage magnitude at each bus, we develop either a linear regression or a fully connected neural network to approximate the inverse AC power flow mappings. The proposed multi-task learning technique further improves the accuracy of branch flow estimation by incorporating the errors of voltage angle differences into the loss function design. Tested on IEEE-300 and IEEE-1354 bus systems with real data, the proposed methods achieve better performance in estimating voltage phasors and branch flows.
\end{abstract}

%---------------------------
\section{Introduction}\label{sect:intro}
Compared with conventional energy sources, solar photovoltaic (PV) generations are more environmentally friendly with almost no greenhouse gas emissions \cite{Chiradeja}. However, the high penetration of PV generations in power systems presents new challenges to grid operators because the solar power outputs are sensitive to ambient conditions. The inherent randomness of PV generation outputs brings significant fluctuations to the state of the system (i.e., voltage phasors). The probabilistic power flow (PPF) study is a key tool in describing the statistical properties of the output variables (i.e., voltage phasors and branch flows)  with respect to the random input variables  (i.e., power injections).

Several approaches have been proposed to solve the PPF problem \cite{Borkowska1974_ppf}. They can be divided into three categories: analytical, approximate, and numerical methods.

\begin{itemize}

    \item \emph{Analytical methods} (e.g., convolution techniques \cite{Allan1981} and cumulant methods \cite{Miao2012}) attempt to obtain the probability density functions (PDFs) of the output variables given the inputs' PDFs. The central concept is that the output variables are represented by a linear combination of input variables based on the first-order Taylor series expansion. This type of methods may lead to significant errors when input variables have large variations. 

    \item \emph{Approximate methods} (e.g., point estimate methods \cite{Juan2007}) aim to approximate moment statistics of output variables based on the  inputs' moments. However, the approximation accuracy is decreased significantly with the increasing number of input variables. Based on the obtained moments, approximate methods work with series expansions to further estimate the needed PDFs. These series expansions  potentially have an issue of convergence guarantees.

    \item \emph{Numerical methods} (e.g., Monte Carlo simulation (MCS) \cite{Mahdi2013}) generate power injections samples and run power flow analysis (using Newton–Raphson method) to calculate the corresponding voltage phasors whose statistical properties can be further obtained. If Newton–Raphson method converges to the true voltage phasors, MCS is the most accurate approach. However, its computation burden can be very heavy for large-scale systems, which is not suitable for real-time applications.

\end{itemize}

The power flow analysis essentially is to calculate the mapping from input variables to output variables. Machine learning approaches use historical data to learn such a mapping, such as radial basis function neural networks \cite{Baghaee2017} and linear regression \cite{Liu2019}. However, shallow neural networks and linear model cannot learn the complex features. \cite{Yang2020} suggests using fully connected neural networks (FCNNs) to approximate the inverse AC power flow (AC-PF) equations to replace numerous iterations in MCS. In their work, the neural networks yield normalized outputs, which may lead to bad estimates of the voltage phasors and fail to accurately capture the statistical properties. In addition, the training phase can be hard since too many different tasks are combined in the loss function. Note that it is often challenging for a single FCNN to adjust the weights for optimizing all tasks \cite{Kendall2018}. Inspired by previous works, the main contribution of this paper is three-fold:

\begin{enumerate}
    \item Voltage magnitudes and angles are two responses with different distributions. We design two different FCNNs to deal with them separately.
    
    \item By using historical data, the load buses with small and large variations in voltage magnitudes are split into two subsets. Based on this characteristic and the linearized power flow analysis, two machine learning techniques including ordinary least-square linear regression and FCNNs are developed.
    
    \item It is clear that branch power flows are directly related to voltage angle differences. In multi-task learning, we design a loss function that incorporates both the voltage angles and angle differences to further improve the accuracy of branch flows estimation.  

\end{enumerate}

The paper is organized as follows. Section \ref{model} presents the mathematical formulation of power flow analysis and describes the PPF problem from the data-driven perspective. Section \ref{proposed} introduces the proposed data-driven method and multi-task learning for estimating voltage phasors. Section \ref{sec:tests} numerically validates the accuracy of proposed methods. Finally, the conclusion is drawn in Section \ref{con}.

%-----------------------
\section{Problem Statement} \label{model}
%-----------------------
\subsection{Power Flow Analysis} 
As a building block of the PPF problem, power flow study numerically analyzes the steady state of a power system (typically in normal operation). In this section, we first introduce the mathematical formulation of the PF study. 

Consider a power network consisting of $N$ buses and $M$ transmission lines. There are three different types of buses: the set of $N_l$ load buses (denoted by $\mathcal{N}_l$), the set of $N_g$ generator buses (denoted by $\mathcal{N}_g$), and one slack bus. Each bus is associated with four quantities, namely active and reactive power injections as well as voltage magnitude and angle. For the PF study, only two quantities are specified while the other two are unknown. Specifically, active and reactive power injections are given for load buses. The active power injections and voltage magnitudes are specified for generator buses. The voltage magnitudes and angles are fixed for the slack bus. In addition, each power line has the active and reactive branch flows as the unknown quantities. 

Based on the law of conservation of energy, the AC-PF model essentially captures the following $2 N_l + N_g$ nodal power balance equations:
\begin{subequations}
\label{AC-PF}
\begin{align}
    P_i &= V_i \sum_{j=1}^N V_j (G_{ij}\cos \theta_{ij} + B_{ij}\sin \theta_{ij} ) \, , i\in \mathcal{N}_l\cup\mathcal{N}_g \, , \\
    Q_i &= V_i \sum_{j=1}^N V_j (G_{ij}\sin \theta_{ij} - B_{ij}\cos \theta_{ij} ) \, , i\in \mathcal{N}_l \, , 
\end{align} 
\end{subequations}
where $P_i$ and $Q_i$ are active and reactive power injections of bus $i$. $V_i$ is the voltage magnitude of bus $i$. $\theta_{ij} := \theta_i - \theta_j$ is the voltage angle difference between bus $i$ and $j$. $G_{ij}$ and $B_{ij}$ are the real and imaginary parts of the $(i,j)$-th element of nodal admittance matrix $\mathbf{Y} \in \mathbb{C}^{N\times N}$, respectively. 

Let $\mathbf{x} = [\mathbf{P}_g; \mathbf{P}_l; \mathbf{Q}_l]$ collect all specified power injections, where $\mathbf{P}_g$, $\mathbf{P}_l$ are active power injections of generator and load buses, and $\mathbf{Q}_l$ collects reactive power injections of load buses. Similarly, the unknown voltage angles and magnitudes are denoted by $\mathbf{y}_a = [\boldsymbol{\theta}_g; \boldsymbol{\theta}_l]$ and $\mathbf{V}_l$, respectively. Let $\mathbf{f}_a(\cdot)$ and $\mathbf{f}_m(\cdot)$ denote the mappings from input variables $\mathbf{x}$ to output variables $\mathbf{y}_a$ and $\mathbf{V}_l$. To this end, the inverse mapping of the AC-PF equations~\eqref{AC-PF} can be compactly rewritten as: 
\begin{subequations}
\label{eq:inversePF}
\begin{align}
    \mathbf{y}_a &= \mathbf{f}_a(\mathbf{x}) , \\
    \mathbf{V}_l &= \mathbf{f}_m(\mathbf{x}) .
\end{align}
\end{subequations}

Upon obtaining all voltage magnitudes and angles,  the branch power flows can be readily computed as follows: 
\begin{subequations}
\label{bf}
\begin{align}
    P_{ij} &= -G_{ij}V_i^2 + V_iV_j(G_{ij}\cos \theta_{ij} + B_{ij}\sin \theta_{ij}) , \\
    Q_{ij} &= B_{ij}V_i^2 + V_iV_j(G_{ij}\sin \theta_{ij} - B_{ij}\cos \theta_{ij}) - \frac{b_{ij}^c}{2}V_i^2, 
\end{align}
\end{subequations}
where $b_{ij}^c$ is the total line-charging susceptance between bus $i$ and bus $j$. 

\subsection{From PF to PPF Analysis}

For MCS-based PPF study, it requires repeatedly solving the inverse AC-PF equations~\eqref{eq:inversePF} for the given power injection samples to obtain the voltage phasors. The overall computational burden is very high due to a tremendous computation involved in the Newton–Raphson method. It is well known that Newton's method is sensitive to the initial point without convergence guarantees. 

Therefore, we are motivated to train a machine learning model to approximate the inverse mapping~\eqref{eq:inversePF} by using historical data. From the data-driven perspective, there are two stages involved: training and inference. In training, by taking in lots of power injection samples, an optimizer adjusts the model's weights to minimize a data fitting loss. The training stage is often time consuming, which however can be done offline. In the inference stage, the testing samples are fed into the trained model whose outputs are the corresponding voltage phasors. The computation time of testing is very low. Therefore, the learned model can serve as the workhorse of the PPF study by surrogating the inverse AC-PF mapping.

\section{Proposed Methods for PPF} \label{proposed}
In this section we propose a variation-cognizant method to approximate the nonlinear inverse power flow mapping $\mathbf{f}_m(\cdot)$, as well as a multi-task learning technique for $\mathbf{f}_a(\cdot)$. 
%-----------------------
\subsection{Variation-cognizant Method for Voltage Magnitudes}
For a per unit system, voltage magnitudes typically vary in a much smaller range than the one for voltage angles. We split the load buses into two disjoint subsets according to the standard deviation (std) values of the voltage magnitudes, which can be estimated by using the historical data: 
\begin{align}
 \mathcal{N}_l = \mathcal{N}_{ls} \cup \mathcal{N}_{lb}~~ \text{and}~~\mathcal{N}_{ls} \cap \mathcal{N}_{lb} = \emptyset,
\end{align}
where subset $\mathcal{N}_{ls} := \{\mathcal{N}_{l}~|~\text{std}(V_l) \leq \gamma\}$ collects load buses with small std values of voltage magnitudes while subset $\mathcal{N}_{lb} := \{\mathcal{N}_{l}~ |~\text{std}(V_l) > \gamma\}$ is for the load buses with big std values. The threshold $\gamma$ can be learnt by a validation process. 

We propose running two different machine learning models for these two subsets. For load buses in $\mathcal{N}_{lb}$, a FCNN will be well trained to approximate the nonlinear mapping from power injections to the voltage magnitudes while capture their big variations. For load buses in $\mathcal{N}_{ls}$, a simple approach is to estimate the voltage phasors by using an ordinary least-squares linear regression. The effectiveness of linearization of the inverse AC-PF mappings has been shown by existing works (e.g., \cite{Liu2019} and \cite{Yang2017}). Consider an extreme case when the std is close to zero, i.e., the voltage magnitude remains almost as a constant regardless of the variation of power injections. In this case, it is reasonable not to employ a complicated FCNNs to avoid possible overfitting and heavy computational burden. 

%-----------------------
\subsection{Multi-task Learning Approach for Branch Flows}
% Compared with single-task learning, multi-task learning optimizes multiple tasks simultaneously, typically using one single neural network. The learning of each task is expected to help the other tasks, and thus the learning efficiency is high. 
Multi-task learning is popular in many real applications, such as computer vision and natural language processing. However, it is faced with a challenge that is increasing the performance of one task may deteriorate the performance of other tasks. Therefore, the issue of which tasks should be learned together and how to balance their importance is critical. In power flow analysis, voltage angles and voltage angles differences are closely related to each other, which naturally motivates us to apply multi-task learning in the neural network training. 

When it comes to estimating voltage angles, previous works only aim to minimize the errors between the estimate and true values, which is a single task. As shown in \eqref{bf}, the branch flows are dictated by the voltage angle differences between connected bus pairs. Hence, we combine voltage angles and angle differences in the FCNN loss function to improve the accuracy of estimating branch flows. 

Given the network topology, voltage angle differences are uniquely determined by the voltage angles, but not vice versa. Thus, more accurate voltage angles estimates don't always show more accurate voltage angles differences estimates. Nevertheless, estimating the voltage and voltage angles difference are two highly related tasks. Hence, we propose a multi-task learning approach by designing the weighted loss function:
\begin{align}
    \mathcal{L}_{\text{new}} &= \mathcal{L}(\Delta{\mathbf{y}_\text{a}}) + \alpha \mathcal{L}({\Delta{\mathbf{y}_{\text{ad}}}})\\
    &= \mathcal{L}(\bar{\mathbf{y}}_\text{a} - \mathbf{y}_\text{a}) + \alpha \mathcal{L}(\bar{\mathbf{y}}_\text{ad} - \mathbf{y}_\text{ad}),
\end{align}
where $\mathcal{L}$ is the mean square error and $\alpha$ is the weighting parameter to balance the losses of the two tasks. True voltage angle differences $\mathbf{y}_\text{ad} = \mathbf{A}\mathbf{y}_a$, where $\mathbf{A} \in \mathbb{R}^{M \times (N-1)}$ is the reduced incidence adjacency matrix (the column related to the slack bus is deleted) whose element is 1 for the \emph{from node} while -1 for the \emph{end node} of a branch. The estimate angle differences $\bar{\mathbf{y}}_\text{ad}$ can be obtained in a similar fashion.

To this end, we compare our proposed approaches (M3 \& M4) with two existing  methods (M1 \& M2) as listed in below:
\begin{itemize}
\item M1: Postulate a linear model 
$\begin{bmatrix}
\mathbf{y}_a \\
\mathbf{V}_{l} 
\end{bmatrix} =  
\mathbf{H}\mathbf{x} + \boldsymbol{\epsilon}$, and solve the ordinary least-squares linear regression problem to estimate both voltage magnitudes and angles. 

\item M2: Train a FCNN to output both voltage magnitudes and angles; see \cite{Yang2020}.

\item M3: Train two separate FCNNs to estimate voltage magnitudes and angles, respectively.

\item M4: Run a data-driven method with multi-task learning to estimate voltage magnitudes and angles, respectively. 
\end{itemize}

\begin{remark}
The proposed M3 is a simple extension of M2. Our motivation is that the underlying characteristics (e.g., value ranges and variations) of voltage magnitudes and angles are quite different. It may be better to train two different neural networks to deal with them separately.  Such a design can generally improve the estimation performance as shown later. The proposed M4 naturally combines the ideas of M1 and M3, where we find that voltage magnitudes with small deviations should be handled by the linear regression. This can further reduce the computational burden and boost the estimation accuracy. 
\end{remark}

% %----------------------
\section{Numerical Results} \label{sec:tests}
In this section, we show the effectiveness of the proposed approaches that are tested on the IEEE-300 and IEEE-1354 bus benchmark systems.
%-------------------------
\subsection{Data Generation and Simulation Setup}
PV power outputs and active power demand of 20 load buses are generated based on the data in \cite{nrel} and \cite{Hong2014}, respectively. Multiplying the active power injections by a uniform random variable in $(0,1)$, we get the corresponding reactive power injections. The active and reactive power demands of all remaining load buses follow a joint multivariate Gaussian distribution. 
The mean values are equal to the original demands. The std values are 0.01 of the mean values for the IEEE-300 bus and 0.1 for the IEEE-1354 bus system, respectively. The correlation coefficients of active and reactive power demands between different buses are 0.2 and 0.8.

Based on the Newton-Raphson algorithm in \texttt{Matpower 7.0} \cite{Daniel2011}, we run MCS to generate the ground true data of voltage phasors. Then, the active and reactive branch flows are computed by using equations \eqref{bf}. Table~\ref{hp} shows the hyperparameters of M2-M4, which are tuned based on the validation errors. We choose rectified linear unit (ReLU) as the activation function. The Adam optimizer is used to train the FCNNs based on Pytorch 1.7.1. The training mini-batch size is 32. For the proposed data-driven method, the threshold $\gamma$ is set to $10^{-3}$ based on the validation procedure. For the proposed multi-task learning method, $\alpha$ is set to 1 and 10 for the IEEE-300 and IEEE-1354 bus systems, respectively.  

% The weights with smaller validation errors are saved and updated during the training process. Then, the best weights will be used to calculate the errors on the test dataset for methods comparison. All the methods are trained and tested five times to alleviate possible randomness.   
\begin{table*}
\caption{The hyperparameters of FCNNs}
\label{hp}
\centering
\begin{tabular}{c|c|c|c|c|c}\hline 
Test system & Dataset size  of [training, validation, testing]  & Method & Voltage  & Number of neurons in each layer & Learning rate  \\\hline
\multirow{5}{*}{IEEE-300} & \multirow{5}{*} {[20000, 5000, 5000]} & \multirow{1}{*}{M2} & angle and magnitude & [530 200 200 200 530] & $1 \times 10^{-4}$ \\ 

\cline{3-6} 

& & \multirow{2}{*}{M3} & angle & [530 300 300 299] & $5 \times 10^{-5}$ \\ 

& & &  magnitude & [530 200 200 231] &  $1 \times  10^{-4}$ \\ 

\cline{3-6} 

& & \multirow{2}{*}{M4} & angle & [530 300 300 299] & $5 \times 10^{-5}$  \\ 

& & &  magnitude & [530 200 200 137] & $1 \times 10^{-4}$ \\ \hline

\multirow{5}{*}{IEEE-1354} & \multirow{5}{*} {[26000, 6000, 6000]} & \multirow{1}{*}{M2} & angle and magnitude & [2447 300 300 300 300 2447] & $7 \times 10^{-5}$ \\

\cline{3-6} 

& & \multirow{2}{*}{M3} & angle & [2447 400 400 400 1353] & $5 \times 10^{-5}$ \\

& & & magnitude & [2447 400 400 400 1094] & $7 \times 10^{-5}$ \\

\cline{3-6}

& & \multirow{2}{*}{M4} & angle & [2447 400 400 400 1353] & $7 \times 10^{-5}$ \\ 

& & & magnitude & [2447 400 400 400 718] & $7 \times 10^{-5}$ \\ 

\hline
\end{tabular}
\end{table*}

\subsection{Performance Metrics}
Let $\bar{\mathbf{O}}$, $\mathbf{O} \in \mathbb{R}^{n \times d}$ denote the estimate and the ground truth of the quantity of interest. Here, $n$ and $d$ are the numbers of testing data points and the responses, respectively. Three performance metrics are used to provide a comprehensive evaluation of different methods.

\begin{itemize}
\item Average root mean square error (RMSE) of all responses:
\begin{align}
   \text{Average RMSE} = \frac{1}{d} \sum_{i=1}^d \sqrt{\frac{1}{n}\|\mathbf{\bar{O}}(:,i)-\mathbf{O}(:,i)\|_2^2},   
\end{align}
where $\mathbf{O}(:,i)$ denotes the $i$-th response (i.e., column of matrix $\mathbf{O}$), and similarly for $\bar{\mathbf{O}}(:,i)$.

\item Average Wasserstein distance \cite{wgan}: 
    \begin{align}
        \text{AWD} = \frac{1}{d} \sum_{i=1}^d \mathcal{W}_1(\bar{\rho}_i, \rho_i), 
    \end{align}
    where $\mathcal{W}_1$ is the first-order Wasserstein distance, and $\bar{\rho}_i$ and $\rho_i$ are the probability distributions of the $i$-th response.

\item Average mean absolute error (MAE) of the mean value $e_1$ and the std value $e_2$:
    \begin{align}
         e_1  = \frac{1}{d}\sum_{i=1}^d | \bar{\mu}_i - \mu_i |, \\
         e_2  = \frac{1}{d}\sum_{i=1}^d | \bar{\sigma}_i - \sigma_i |,
    \end{align}
    where $\bar{\mu}_i$ and $\mu_i$ are the mean values of the $i$-th response, and $\bar{\sigma}_i$ and $\sigma_i$ are the corresponding std values. 
\end{itemize}

\subsection{Test Results}
% We applied the aforementioned four different methods to estimate voltage magnitudes of load buses, as well as voltage angles of all buses except for the slack bus.
Table~\ref{rmse_mag} shows the voltage magnitude estimation errors tested on the two benchmark systems. Clearly, with the combination of linear regression and FCNN, the proposed data-driven method achieves the best performance. 

As shown in Tables \ref{rmse_ang} and \ref{rmse_branch}, the proposed multi-task learning method improves the accuracy of angle difference estimation. Thus, its branch flow estimations are the most accurate among all methods for different cases. For the IEEE-1354 bus system, the multi-task learning method yields slightly worse estimates of voltage angles, compared with approach M3. However, it performs better in the angle difference estimates, which yields better estimates of the branch power flows. It is worth pointing out that for the multi-task learning best performance of one task cannot guarantee the performance of the other task. However, it turns out that improving the estimation of voltage angles differences does not affect too much the accuracy of estimating voltage angles themselves since these two tasks are highly related.  

\begin{figure}[t]
    \centering
    \includegraphics[scale=0.55]{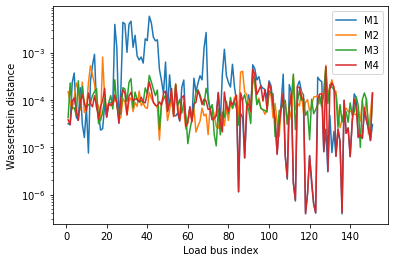}% second figure itself
    \caption{The Wasserstein distance of voltage magnitude distributions for IEEE-300 bus system.}
    \label{WD_mag_300}
\end{figure}

\begin{figure}[t]
    \centering
    \includegraphics[scale=0.55]{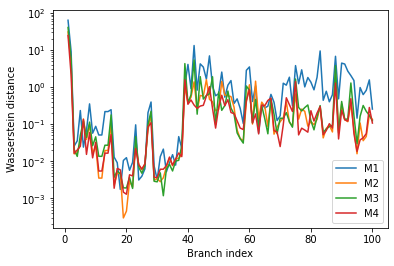}% second figure itself
    \caption{The Wasserstein distance of active branch flow distributions for IEEE-300 bus system.}
    \label{WD_ABF_300}
\end{figure}

\begin{figure}[t]
    \centering
    \includegraphics[scale=0.55]{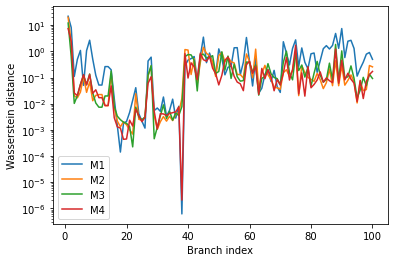}% second figure itself
    \caption{The Wasserstein distance of reactive branch flow distributions for IEEE-300 bus system.}
    \label{WD_RBF_300}
\end{figure}

As shown in Table \ref{AWD_vol} and Fig.\ref{WD_mag_300}, the data-driven method has the least AWD of voltage magnitudes. In addition, Table \ref{AWD_BF}, Fig.\ref{WD_ABF_300} and Fig.\ref{WD_RBF_300} show the proposed M4 achieves the best performance in branch flows distribution estimates. 

In addition to AWD, we also compare different methods in terms of the mean and std values of the estimates, as shown in Table \ref{mean_std}. For reactive branch flows, the proposed M4 has the smallest MAEs. For active branch flows, M4 reduces the MAEs of the std at least by half, compared with the other three alternatives. 

% The MAE of std value estimates of the  is around two times better than M2 and three times better than M3 on the IEEE-300 bus system and is about two times better than M2 and M3 on the IEEE-1354 bus system. 

\begin{table}
\caption{Average RMSEs of voltage magnitudes  ($10^{-4}$)}
\label{rmse_mag}
\centering
\begin{tabular}{c|c|c|c|c}\hline 
Test system & M1 & M2 & M3 & M4 \\ \hline
\multirow{1}{*}{IEEE-300}  & 16.15 & 4.36 & 3.68 & \textbf{3.34} \\ \hline
\multirow{1}{*}{IEEE-1354} & 2.58 & 6.81 & 2.75 & \textbf{2.11} \\ \hline
\end{tabular}
\end{table}

\begin{table}
\caption{Average RMSEs of voltage angles and angles differences calculations  ($10^{-3}$)}
\label{rmse_ang}
\centering
\begin{tabular}{c|c|c|c|c|c}\hline 
Test system & Voltage & M1 & M2 & M3 & M4 \\\hline
\multirow{2}{*}{IEEE-300} & angle & 23.15 & 2.31 & 2.24 & \textbf{2.10}\\
& angle difference & 0.91 & 0.51 & 0.50 & \textbf{0.38} \\ \hline
\multirow{2}{*}{IEEE-1354} & angle & 5.96 & 1.53 & \textbf{0.97} & 1.11\\
& angle difference & 0.84 & 0.81 & 0.59 & \textbf{0.53}\\ \hline
\end{tabular}
\end{table}

\begin{table}
\caption{Average RMSEs of branch flows calculations }
\label{rmse_branch}
\centering
\begin{tabular}{c|c|c|c|c|c}\hline 
Test system & Branch flow & M1 & M2 & M3 & M4  \\\hline
\multirow{2}{*}{IEEE-300} & active & 2.92 & 1.51 & 1.69 & \textbf{0.93} \\
& reactive & 1.23 & 0.96 & 0.70 & \textbf{0.56} \\ \hline
\multirow{2}{*}{IEEE-1354} & active &  22.12 & 8.67 & 7.89 & \textbf{5.60} \\
& reactive & 5.51 & 5.55 & 3.01 & \textbf{2.03}\\ \hline
\end{tabular}
\end{table}

\begin{table}
\caption{AWD of the voltage phasors distributions  ($10^{-4})$}
\label{AWD_vol}
\centering
\begin{tabular}{c|c|c|c|c|c}\hline 
Test system & Voltage & M1 & M2 & M3 & M4  \\\hline
\multirow{2}{*}{IEEE-300} & angle & 139.544 & 5.580  & 5.104 &  \textbf{5.102}  \\
& magnitude & 8.136 & 1.057  & 1.182 & \textbf{0.917}   \\ \hline
\multirow{2}{*}{IEEE-1354} & angle & 25.199 & 3.840  & \textbf{3.066} & 3.629 \\
& magnitude & 0.969 & 1.548 & 0.946 & \textbf{0.674}  \\ \hline
\end{tabular}
\end{table}

\begin{table}
\caption{AWD of the branch flows distributions }
\label{AWD_BF}
\centering
\begin{tabular}{c|c|c|c|c|c}\hline 
Test system & Branch flow & M1 & M2 & M3 & M4  \\\hline
\multirow{2}{*}{IEEE-300} & active & 1.843 & 0.527  & 0.772 &  \textbf{0.403}  \\
& reactive & 0.579 & 0.369  & 0.323 & \textbf{0.228}   \\ \hline
\multirow{2}{*}{IEEE-1354} & active & 16.511 & 3.975  & 3.853 & \textbf{2.788} \\
& reactive & 3.979 & 2.998 & 1.827 & \textbf{1.177}  \\ \hline
\end{tabular}
\end{table}

\begin{table}[t]
\caption{Average MAE of estimate mean and std of branch flows }
\label{mean_std}
\centering
\begin{tabular}{c|c|c|c|c|c|c}\hline 
Test system & Branch flow & & M1 & M2 & M3 & M4  \\\hline
\multirow{4}{*}{IEEE-300} & \multirow{2}{*}{active} & mean & 1.27  & \textbf{0.16} & 0.38 & 0.29  \\ 
& & std & 1.27 & 0.57  & 0.72 & \textbf{0.28} \\ 
\cline{2-7}
& \multirow{2}{*}{reactive} & mean & 0.22 & 0.19  & 0.23 & \textbf{0.14}   \\ 
& & std & 0.36 & 0.34  & 0.19 & \textbf{0.11}  \\ \hline
\multirow{4}{*}{IEEE-1354} & \multirow{2}{*}{active} & mean & 12.43 & \textbf{1.79}  & \textbf{1.79} & 1.95  \\
& & std & 12.98 & 4.30 & 4.13 & \textbf{2.09}\\
\cline{2-7}
& \multirow{2}{*}{reactive} & mean & 2.88 & 1.75  & 1.32 & \textbf{0.94}   \\ 
& & std & 3.11 & 3.30  & 1.37 & \textbf{0.70}  \\ \hline
\end{tabular}
\end{table}

\section{Conclusion} \label{con}
A straightforward approach of PPF analysis is to repeatedly compute voltage phasors for each sample of the given power injections. Such a Monte Carlo simulation is computationally heavy that is not suitable for large-scale systems and real-time applications. 
To reduce the computational burden, we propose an efficient data-driven method alongside multi-task learning to approximate the mappings from power injections to voltage phasors. Based on historical data, the load buses are split into two subsets that are deal with by two different models. For multi-task learning, our meticulous design of the weighted loss improves the estimation accuracy for power line flows. Tested on two IEEE bus systems with synthetic and real data, the simulation results corroborate the effectiveness of the proposed methods in estimating voltage phasors and branch power flows.

\nocite{*}
\bibliographystyle{IEEEtran}
\bibliography{PFrefs,IEEEabrv}

\end{document}